\documentclass[journal]{IEEEtran}

\IEEEoverridecommandlockouts

\usepackage{amsmath,amssymb,amsfonts}
\usepackage{graphicx}
\usepackage{textcomp}
\usepackage{relsize}

\usepackage{color}
\usepackage{pifont}
\usepackage{url}
\usepackage{cite}

\newtheorem{mydef}{Definition}
\usepackage{mathtools}
\usepackage{multirow}
\usepackage{comment}
\usepackage{booktabs}
\usepackage{algorithm}
\usepackage{subfigure}
\usepackage{algpseudocode}
\usepackage{listings}

\usepackage{placeins}

\usepackage{lscape}

\usepackage{tikz,pgfplots}
\usepackage{pgfplotstable}
\usepackage{pgf-pie}

\pgfplotsset{width=7cm,compat=1.3}
\usetikzlibrary{patterns}

\usetikzlibrary{arrows}
\usetikzlibrary{shapes}

\usepackage{textcomp}

\def\BibTeX{{\rm B\kern-.05em{\sc i\kern-.025em b}\kern-.08em
    T\kern-.1667em\lower.7ex\hbox{E}\kern-.125emX}}

\ifCLASSOPTIONcompsoc
  \usepackage[nocompress]{cite}
\else
  \usepackage{cite}
\fi

\ifCLASSINFOpdf
  
\else
 
\fi
\begin{document}

\title{Threat-Specific Risk Assessment for IP Multimedia Subsystem Networks Based on Hierarchical Models\\
}



\author{\IEEEauthorblockN{Abdullah Ehsan Shaikh, Simon Yusuf Enoch}\\
\IEEEauthorblockA{ School of Information Technology,\\
Whitecliffe College, \\New Zealand.\\
20231074@mywhitecliffe.com, https://orcid.org/0000-0002-0970-3621}
}

\maketitle

\begin{abstract}
Over the years, IP Multimedia Subsystems (IMS) networks have become increasingly critical as they form the backbone of modern telecommunications, enabling the integration of multimedia services such as voice, video, and messaging over IP-based infrastructures and next-generation networks. However, this integration has led to an increase in the attack surface of the IMS network, making it more prone to various forms of cyber threats and attacks, including Denial of Service (DoS) attacks, SIP-based attacks, unauthorized access, \textit{etc}. As a result, it is important to find a way to manage and assess the security of IMS networks, but there is a lack of a systematic approach to managing the identification of vulnerabilities and threats. In this paper, we propose a model and a threat-specific risk security modeling and assessment approach to model and assess the threats of the IMS network. This model will provide a structured methodology for representing and analyzing threats and attack scenarios in layers within a hierarchical model. The proposed model aims to enhance the security posture of IMS networks by improving vulnerability management, risk evaluation, and defense evaluation against cyber threats.
We perform a preliminary evaluation based on vulnerability collected from the National Vulnerability Database for devices in the IMS network. The results showed that we can model and assess the threats of IMS networks. IMS network defenders can use this model to understand their security postures taking into account the threat and risk posed by each vulnerability.
\end{abstract}


\IEEEpeerreviewmaketitle

\section{Introduction}
\label{sec:intro}

The IP Multimedia Subsystem (IMS) is a critical architectural framework in modern telecommunications, enabling the seamless integration of multimedia services such as voice, video, and messaging over IP networks. As IMS continues to evolve, it has become a cornerstone of Next-Generation Networks (NGN), facilitating a wide range of applications across both mobile and fixed network environments. However, the open and distributed nature of IMS introduces significant security threats such as denial of service (DoS), spoofing, unauthorized access, \textit{etc}, primarily due to the vulnerabilities inherent in IP-based systems that affect both network providers and network users \cite{wang2009model}. Moreover, different functions within the IMS are susceptible to different types of security threats, each requiring specific security countermeasures. Hence, it is important to assess the different threats posed by each vulnerability in the IMS network in order to effectively defend against them.

Several organizations have made recommendations on how to improve the security of IMS networks. For example, the 3rd Generation Partnership Project (3GPP) outlined how to develop a robust security framework for IMS networks addressing authentication, encryption, and integrity protection to ensure secure communication and data protection in the IMS network \cite{damasceno2024open}. The ITU-X recommendations cover the issues of information security standards, ensuring the interoperability, security, and scalability of global communication networks \cite{liu2024enhancing}. However, the recommendations and standards presented are focused on end-to-end security policies and governance. 

Threat Vulnerability and Risk Analysis (TVRA) model is used to identify risk to the system based on the probability of attack success and attack impact on the system \cite{rizzo2011etsi}. The TVRA is primarily used to justify the development of security countermeasures in a system rather than understanding the security posture of the IMS network. Besides, it has failed to provide a practical method for collecting and assessing the IMS system's vulnerabilities. 

Automated security modeling and assessments (e.g., Attack Trees) can help collect system vulnerabilities, and threats, and evaluate defense strategies before deployment. However, the current risk modeling approach used in enterprise networks, cloud, or IoT cannot be used in its current form because it cannot capture the unique characteristics of the IMS system settings and networks. In addition, most existing model-based security assessments are theoretically based or end-to-end security approaches \cite{maachaoui2011model,rizzo2011etsi,chen2008efficient} rather than the IMS network modeling and quantitative assessment.

To address this problem, we propose a graphical security model called the IMS-HAG. The IMS-HAG is designed to assess and evaluate the threats and vulnerabilities of the IP multimedia subsystem network. The IMS-HAG systematically identifies potential attack paths that could compromise a target system by considering vulnerabilities and threats. It offers a structured approach to understanding how different vulnerabilities can be exploited to reach and compromise critical systems. Using this model, we can evaluate the vulnerabilities and threats of IMS networks, and also compute threat-specific risk based on different attack scenarios. 

Based on the review of existing studies,  this is the first paper that models and systematically assesses the threat-specific risk of devices on an IMS network. The key contributions of this work are summarized as follows:

\begin{itemize}
 \item Propose a hierarchical graph-based model to model and assess threats to IMS networks
 \item Assess the threats of the IMS network using threat-specific security metrics and attack success probability metrics.
 \item Demonstrate the suitability of the model using real-world data in a simulation.

\end{itemize}
The remaining part of this paper is organized as follows. The related work is presented in Section~\ref{sec:related}. The details of the proposed hierarchical modeling, including its description, formalism, and security metrics are presented in Section~\ref{sec:proposed_system}. Section~\ref{sec:net_attacker_models} presents the IMS core network and the attacker model. Section~\ref{sec:analysis} provides the numerical results from the simulation, and finally, Section~\ref{sec:conclusion} concludes the paper.

\section{Related Work }
\label{sec:related}

In this section, we survey the state-of-the-art literature on modeling and assessing the security of IMS.

Wang and Liu \cite{wang2009model} presented a vulnerability analysis system for IMS networks using simulation, focusing on identifying key security weaknesses. Ed-daoui \textit{et al.} \cite{ed2016security} introduced a layered and tiered architectural model to enhance IMS security without impacting performance. The authors proposed a practical framework for improving IMS security, potentially reducing vulnerabilities to specific types of attacks. However, the system was based on a vulnerability risk metric and did not capture the threats posed to the system.

Belmekki \textit{et al.} \cite{belmekki2013towards} developed a system to reduce the probability of attack success on critical IMS interfaces. The system uses encryption and authentication to mitigate unauthorized access and tampering. Sisalem \textit{et al.} \cite{sisalem2006denial} created a security framework for SIP-based Denial of Service (DoS) attacks in IMS. The authors conducted a scenario-based analysis of SIP-DoS attacks and validated their model in real-world SIP environments. Their approach focuses solely on SIP functions and does not address other types of attacks. Chen \cite{chen2006detecting} proposed detection techniques for SIP-based DoS attacks, but this approach is similarly limited to specific types of DoS attacks and does not cover other attack scenarios. Geneiatakis \textit{et al.} \cite{geneiatakis2005framework} presented a framework for detecting malformed messages in SIP networks to enhance security. However, this approach remains a theoretical framework without extensive real-world testing or validation.

McGann \textit{et al.} \cite{mcgann2005analysis} provided an analysis of security threats and tools in SIP-based VoIP systems, focusing on vulnerabilities and countermeasures. The purpose of their system is to guide the development of more effective security tools for SIP-based VoIP systems. Sengar \textit{et al.} \cite{sengar2006voip} introduced an intrusion detection approach for VoIP using interacting protocol state machines, along with the analysis of attack scenarios. Lin \textit{et al.} \cite{lin2009quality} presented a Queuing Petri Net model to assess the IMS performance, focusing on SIP signaling. The authors defined seven IMS security levels based on 3GPP standards and introduced a Quality of Protection partition model reflecting SIP protection and user security needs, including a method to quantitatively assess the impact of security mechanisms on system performance costs, and the balance between security and performance.
Mauro \textit{et al.} \cite{9439188} modeled resource allocation for softIMS nodes as a type of queueing model and used Stochastic Reward Net to assess availability. The authors introduced two routines, OptCNT and OptSearchChain, for performance and availability analysis. The focus of the existing literature differs from our approach. Our focus is to develop a hierarchical, graph-based model to assess a threat-guided risk (of all threats or sub-threats) for functions found in the IMS network.

\section{A Proposed IP Multimedia Subsystem Networks Security Model}
\label{sec:proposed_system}

This section will provide the description and the formal definition of the proposed IMS-HAG, including the threat-specific risk assessment approach for the threat evaluation. Then, the approach to collect and estimate IMS network vulnerabilities and threats follows.

\subsection{IP Multimedia Subsystem Networks}
\label{sec:imsHAG}

IMS provides a standardized platform for delivering Internet services over IP (e.g., delivering data, voice, video, messaging, and session management. So, the key components/functions of the IMS include; Call Session Control Functions (CSCF), Media Resource Functions (MRF), Home Subscriber Server (HSS), Application Servers (AS), Media Gateway Control Function (MGCF), Session Border Controller (SBC), Subscriber Location Function (SLF), Breakout Gateway Control Function (BGCF), Media Resource Function Processor (MRFP), etc.

So, the IMS system consists of various sub-systems; some of which are accessible to external users, while others are designated for internal operations.  Our scope is to model and assess the threats of functions that are open to external users on the IMS network (i.e., the edge devices that serve as interfaces between the external user and the IMS network). Therefore, we propose a threat security model for the IMS network named {\em IMS-HAG}, based on the work in Hierarchical Attack Representation Model~\cite{hong2016towards}. 
The IMS-HAG is planned to capture the functions running on IMS systems, including their connections, vulnerabilities, and events. The system functions are captured as nodes and the connections between the functions as edges. Also, logical gates are used to model single or multiple attack goals. For the vulnerabilities, threats, and impact values are identified based on possible vulnerability threats as defined in NVD \cite{nvd} and then incorporated into the model.

In the following, we define the specifications and features of the IMS-HAG:

\begin{itemize}

\item System subnet; a collection of systems that perform similar tasks (e.g., several systems for Media Resource Functions). 

\item Functions; has unique names and  IP addresses. Each function is a system.

\item IMS configuration; the set of rules that are associated with routing, open ports, etc. This can be collected using NMap, OpenVAS~\cite{openvas} or provided by the network administrator. 

\item  Function vulnerabilities and metrics; can be collected using automated scanners such as Nessus~\cite{nessus}. Each vulnerability has a metric (e.g., Common Vulnerability Scoring System (CVSS) Base Score, privilege level (e.g., user or admin), and a set of threats posed based on the vulnerability, such threats including information disclosure, tampering, denial of service, escalation of privileges, \textit{etc}~\cite{stride_model}).

\item topology information (edges); this includes information from the administrator's network map or tools such as NMAP.
\end{itemize}

So, the IMS-HAG is a hierarchical graph-based model where various security information is captured in different layers. Specifically, in the top layer, the functions, topology information, and the attacker's entry points are captured. In the lower layer, the vulnerability and threat information are captured using attack trees. The IMS hierarchical graph-based model (IMS-HAG) is formally represented as follows.

\begin{mydef} \textbf{IMS-HAG} \end{mydef}
The IMS-HAG  captures all potential known attack paths for different attack goals. It is formally defined as a 3-tuple IMS-HAG=$(T, L, M)$, where $T$ stands for the elements in the top layer, $L$ stand for all the elements in the lower layer, such that each node elements from the top layer have a unique attack tree for it. The $M$ stands for the mapping of node elements from the top layer with a unique attack tree in the lower layer.

\begin{mydef} \textbf{The Top Layer} \end{mydef}
The top layer is a graph that models the nodes' topology and attack goals. The top layer is defined as a 4-tuple $U=(N,E, G, C)$, where $N = \{n_1, n_2, \ldots\}$ represents the set of nodes (i.e., for functions) in the top layer, $E = \{e_1, e_2, \ldots\}$ are the connections represented as edges that connect the nodes locate in the top layer (such that $E{\subseteq}N{\times}N$), and $G = \{g_1, g_2, \ldots\}$ represents the different goals (single or multiple) the attacker wants to compromise, such that $T_g {\in}N$. The conditions for combining the attack goals are represented using $C = \{AND, OR,\ldots\} $.

\begin{mydef} \textbf{The Lower Layer} \end{mydef}
Each node from the top layer has an attack tree (AT) in the lower layer. So, the lower layer $L = \{at_{n_1}, at_{n_2}, \ldots\}$, where $at_{n_j}$ is a separate AT linked to a node $n_j$ from the top layer, and node $n_j$ is a 5-tuple $at_{n_j}=(V, GT, f, type, root)$, where $V = \{v_0, v_1, \ldots\}$ is a set of vulnerabilities (such that $v_i$ has a different metrics and a set of STRIDE threats ($TH$)), $GT = \{gt_1, gt_2, \ldots\}$ represent the logical gates that combine the vulnerabilities of a node, $f$ is a function that maps gates to vulnerabilities and other gates, defined as $f\subseteq \{gt_i \rightarrow v_i\}, v_i\in V {\cup}GT$, $type$ defines the type of each gate $gt_i\subseteq \{GT {\rightarrow}\{AND, OR\}\}$. Additionally, $root$ denotes the root privilege of the $at_{n_j}$.

\subsection{Threat-Specific Risk Metric}
\label{secsub:metrics}

Vulnerabilities found on IMS functions pose different types of threats to the network. To provide appropriate defense for these threats, a cybersecurity defender must understand the threats and risks they pose. In this section, we define the threat-specific risk metric, and in Table \ref{tbl:notations}, we provide the notations used in the equations for the threat metrics.

Model-based threat-specific risk can be used to assess threat risk at different levels;  node level,  attack path level, and network level. Below, we provide the formal definition of the Model-based threat-specific risk as used in the security model. The node-level threat risk is calculated based on the threat information of an attack tree belonging to a node. The attack path threat risk is calculated based on the sequence of nodes to the final target for an attack path of interest, while the network-level threat risk is calculated based on all sequences of nodes to the final target in the top layer of the model.  Table \ref{tbl:notations} provides the notations used in the equations for the threat metrics.

\begin{table}[ht]
\centering
\scriptsize
\caption{Notations, functions and their definitions}
\label{tbl:notations}
\begin{tabular}{|ll|}
\hline
\textbf{Symbols} & \textbf{Definition} \\ \hline  \hline
$n_j$ & represent a node \\ \hline
$v_i$ & represent a vulnerability \\ \hline
$path_i$ & represent an attack path  \\ \hline
$p(v_i)$ & function for the prob. of attack success of $v_i$ \\ \hline
$r(root_i)$ & function for the value of the risk of the root of AT \\ \hline
$r(n_j)$ & compute the risk of $(n_j)$ \\ \hline
$aim(v_i)$ & compute the impact value $v_i$ \\ \hline
$aim(n_j)$ & compute the impact of $n_j$ \\ \hline
$th$ & represent a single threat \\ \hline
$th_{v_i, n_j}$ & represent the metric linked to $th$ of $v_i$ on $n_j$ \\ \hline
$V(n_j)$ & represent the set of $V$ on $n_j$ \\ \hline
$TH$ & represent all STRIDE threat category \\ \hline
ES & represent the exploitability score \\ \hline
$Q$ & represent the subset of threats, $Q \subseteq TH$ \\ \hline
$TH(v_i)$ & is all the STRIDE threat of $v_i$ \\ \hline
$Q(v_i)$ & represent the subset of STRIDE threats linked to $v_i$ \\ \hline
$r_{th}(n_j)$ & calculate the risk of $(n_j)$ with all $TH$ \\ \hline
$r_{th}(path_i)$ & calculate the threat risk of $path_i$ \\ \hline
$Risk_{th}$ & calculate the network level threat specific risk \\ \hline
$Risk_{th, Q}$ & represent the subset $Q$ threat-specific risk \\ \hline
$r_{th}(n_j, Q)$ & calculate the risk of a node based on $Risk_{th, Q}$ \\ \hline
$\mathbb{AP}$ & represent the set of paths to attack goal, $\mathbb{AP} = {path_1, \dots}$ \\ \hline
\end{tabular}
\end{table}

\subsubsection{Metric: Node, Attack path, and Network level}
\label{secsub:threat_risk}
This section explains the model-based threat-specific risk metric utilized in this paper. This risk metric quantifies the expected impact on the system, defined as the product of the overall attack impact and the probability of successful attacks.

To account for the various categories of threats, we adopt the concept of threat-specific risk as proposed by ~\cite{nhlabatsi2018threat}, and incorporate it into IMS-HAG. In the calculations, the vulnerability level risk is computed by Equation \eqref{eq_r}, and the node level risk by \eqref{eq_r2}.
Then, the values for the threat-specific risk for the node, attack path, and network level are given by Equation~\eqref{eq_threat1}, \eqref{eq_threat2} and \eqref{eq_threat3} (all threats), respectively.

\begin{equation} \label{eq_r}
r(root_i)=
\left\{
\begin{array}{ll}
\mathlarger \sum\limits_{v_i \in f(gt_i)}{p(v_i)\times aim(v_i)}, & gt_i \in GT \atop f(gt_i)=AND\\
\mathop {max} \limits_{v_i \in f(gt_i)}{p(v_i)\times aim(v_i)}, & gt_i \in GT \atop f(gt_i)=OR
\end{array}
\right.
\end{equation}

\small
\begin{equation} \label{eq_r2}
r(n_j)=r(root_i)
\end{equation}

\begin{equation} \label{eq_threat1}
\begin{aligned}
r_{th}(n_j)={} & {r(n_j) \times \displaystyle \sum th_{v_i, n_j}},  ~\forall~th\in v_i,~v_i~\in V(n_j)
\end{aligned}
\end{equation}

\begin{equation} \label{eq_threat2}
\begin{array}{ll}
r_{th}(path_i)=\mathlarger \sum\limits~{r_{th}(n_j)}, & \forall~n_j\in path_i
\end{array}
\end{equation}

\begin{equation} \label{eq_threat3}
\begin{array}{ll}
Risk_{th}=\mathlarger \sum\limits~{r_{th}(path_i)}, & \forall~path_i\in \mathbb{AP}
\end{array}
\end{equation}

We can compute the threat-specific risk of a function based on a subset of threats, examining how different types of threats contribute to the overall risk of that function/system. We compute the risk linked with a subset of threats on a single node or function using equation~\eqref{eq_threat4}. The path-level risk is computed using equation~\eqref{eq_threat5}, while the network-level risk is computed with equation~\eqref{eq_threat6}.

\begin{equation} \label{eq_threat4}
\begin{aligned}
r_{th}(n_j, Q)={} & {r(n_j) \times \displaystyle \sum_{i=1}^{|Q(v_i)|} th_{v_i, n_j}},  ~\forall~th\in Q, ~v_i\in V(n_j)
\end{aligned}
\end{equation}

\begin{equation} \label{eq_threat5}
\begin{array}{ll}
r_{th}(path_i, Q)=\mathlarger \sum\limits~{r_{th}(n_j, Q)}, & \forall~n_j\in path_i
\end{array}
\end{equation}

\begin{equation} \label{eq_threat6}
\begin{array}{ll}
Risk_{th, Q}=\mathlarger \sum\limits~{r_{th}(path_i, Q)}, & \forall~path_i\in \mathbb{AP}
\end{array}
\end{equation}

\subsubsection{Multiple Functions as Attack Goals}
\label{secsub:multiple_function}

Multiple functions can be targeted by an attacker and this could be on the same attack paths or different attack paths. To compute the threat-specific risk for multiple functions as the final targets, logical gates are used to capture this scenario. 
The threat-specific risk for the attack paths will need to be calculated by Equation~\eqref{eq_threat2}. Then, threat-specific risk when an attacker compromises a single function or multiple functions as the attack goals are described in  Table~\ref{tbl:mult-_goals}.

\begin{table}[ht]
\centering
\caption{Computing multiple targets (goals)}
\label{tbl:mult-_goals}
\begin{tabular}{ll}
\hline
\textbf{Gate type} & \textbf{Formulas} \\ \hline
&Threat Risk\\ \hline
OR gate & $\mathop {max} \limits_{tg_i \in Tg}{tg_{i}}$  \\ \hline
AND gate &  $\displaystyle \sum_{i=1}^{|Tg|}tg_i,  ~tg_i \in Tg$\\ \hline
\end{tabular}
\end{table}

\section{Network Model, Attacker Model \& Metrics Computations }
\label{sec:net_attacker_models}
In this section, the network, attacker, and defense model used in this paper is described.

\subsection{Network model}
\label{sec:netmodel}

We consider a typical connected IMS network that delivers a range of multimedia services over IP-based networks. However, our focus is on the edge devices (e.g., Proxy CSCF (P-CSCF), Session Initiation Protocol Application Proxy (SIP-AS), etc ) that accept outside traffic from external users to the core IMS. We assume these devices have vulnerable applications that can handle the signal, session management, user authentication, and service execution. We assume that the edge devices are located in different subnets with each subnet hosting several functions that provide the interface between the external user and the IMS core. The SBC, in particular, is crucial for managing media streams and controlling traffic flow, while the CSCF forwards SIP messages to the appropriate internal CSCF components for further processing.

\begin{figure}[!htb]
 	\centering
 \includegraphics[width=0.85 \linewidth]{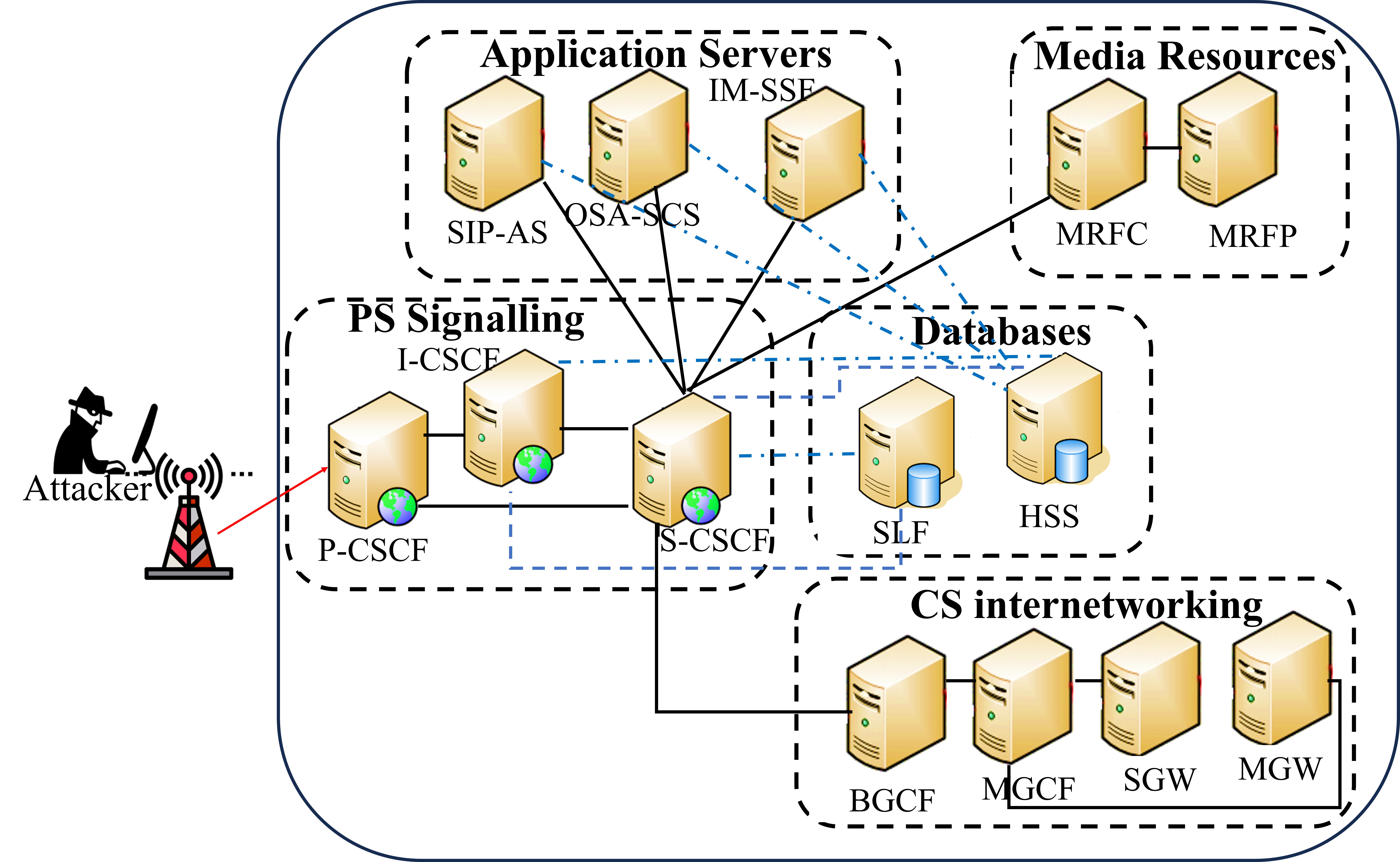}
 	\caption{The IMS network}
 	\label{fig:network}
 \end{figure}

\subsection{Attacker Model}
\label{secsub:attacker}
We assume an external attack can access the network and interact with  VOLTE signaling and wifi access.
We assume the following stages for the targeted cyber-attacks in the attack model. 
\begin{enumerate}
    \item {\em Footprinting and Reconnaissance:} The adversary can gather information about a target organization by using tools such as scanning open ports, mapping network topologies, and collecting details about functions, their operating systems, and IP addresses—to identify potential vulnerabilities.
    \item {\em Exploitation:} Using the vulnerability collected, the adversary can deploy an exploit and exploit it on the target system. 
    \item {\em Command and control:} Once the adversary exploits the vulnerability, the adversary is in command on each side of the system connections and can easily control the system. The adversaries can begin extracting sensitive information and sending them to its end.
    
\end{enumerate}

\subsection{Defense Model}
\label{secsub:defense}

To demonstrate the applicability of the IMS-HAG for evaluating the effectiveness of countermeasures, we considered conventional defense mechanisms such as host isolation to protect against threats and vulnerability patching on critical functions or hosts to fix security weaknesses. We assumed that, due to the availability and importance of the functions, defense cannot be deployed on all systems at the same time. In the IMS-HAG, deploying a defense mechanism can change the structure of attack trees in the lower layers (e.g., when a vulnerability is patched) or the structure of the top layer when a function is isolated or a firewall rule is changed. Since not all defenses can be deployed at once, the functions with the most critical vulnerabilities are patched or isolated first based on threat evaluation. For example - several threats may affect the P-CSCF, but Denial of Service (DoS) is particularly critical due to its role in handling signaling traffic. Therefore, a defender would want to use threat-specific risk assessment using the IMS-HAG, focusing on DoS as a sub-threat to evaluate the effectiveness of defense mechanisms. This evaluation can be based on DoS alone or in combination with other threats, depending on security requirements at that time.

\subsection{Building IMS-HAG with network configurations; topology, vulnerabilities, threats and their metrics}
\label{secsub:asigned_metrics}

To construct the IMS-HAG for security analysis, topology, vulnerabilities, threats, and their metrics must be provided as input. In this paper, the administrator's network map will be used as the topology settings (e.g., hosts IP addresses/names, their reachability information) to build the top layer (other automated network map tools can also be used here), and Nessus scanner will be used to scan and collect vulnerabilities information for each function in the top layer to build the respective ATs for each function in the lower layer. The information collected from Nessus. We collected and used the vulnerability information from the National Vulnerability Database(NVD). We used CVSS Base Score's version 3 - severity and metrics; CVSS Base score, attack impact metric, exploitability score (ES), and a collection of threats linked to it will be captured into the ATs of the node. In this work, we use the CVSS impact value and ES value as attack impact $aim(v_i)$ value and the attack success probability $p(v_i)$ value, respectively.  We normalize the ES for the probability of attack success to the range 0 - 1 using $p(v_i)= (ES/10)$, and then set the $p(v_i)$ for the vulnerability (i.e., [0,1]: with 0 meaning hard to exploit, and 1 indicates that the system is easily exploitable).

For the threats, we use the STRIDE model to identify each threat posed by each vulnerability. The STRIDE threats model categorized threats into six categories ``\textbf{S}poofing, \textbf{T}ampering, \textbf{R}epudiation, \textbf{I}nformation disclosure, \textbf{D}enial of Service, and \textbf{E}levation of privilege''. Based on the description of vulnerabilities provided by MITRE cooperation where threat information such as a vulnerability can cause total I, E, allow information change, \textit{etc}, the model is populated with the vulnerability threat information. 
Also, based on the network administrator's perceived impact of a threat and based on the knowledge of the admin the weight of the impact will be provided as input within the range [0:1], with 1 being the highest threat impact such that the total weight for each vulnerability must be 1. This information is then captured in the AT of the security model and used in the evaluation. 

\begin{table}[!ht]
\centering
\scriptsize
\caption{Threat Weighting Guide}
\label{tbl:weightThreats}
\begin{tabular}{|c|c|c|c|c|c|}
\hline
impact &None& Low risk & Medium risk &High risk & Critical \\ \hline
 weight & 0 & 0.1 - 0.3&0.31 - 0.50&0.51 - 0.80&0.81 - 1.0 \\ \hline
\end{tabular}
\end{table} 

\section{Numerical Analysis and Evaluation}
\label{sec:analysis}

We conduct an experimental analysis through simulations using a sample network. The threat-specific risk metric is employed to evaluate the threats present in the IMS network. For measuring the probability of attack success, we utilize the metrics defined by the authors in \cite{enoch2020composite} for our evaluation.

The focus of the simulation is; (i) model and assess the security of the IMS network using IMS-HAG, (ii) calculate the threat-specific risk at the node level, attack path level, and network level, and (iii) determine the threat-specific risk for sub-systems. Figure \ref{fig:network} illustrates the IMS network utilized in the simulation.

We consider two different attack scenarios for the security evaluation, in order to show the threat related to attacking the different systems. The scenarios are described as follows:

\begin{itemize}
    \item {\em Scenario 1:} An external attacker trying to exploit multiple functions to reach the SIP Application function in the IMS network. The attacker's goal is to exploit the vulnerability (CVE-2018-10544 \cite{CVE-2019-7285}) found on the SIP application by launching a man-in-the-middle attack and modifying the SIP headers or body. They could redirect calls or alter media streams by manipulating the Session Description Protocol within the SIP message body. However, the attacker must go through the P-CSCF and then to the S-CSCF before reaching SIP-AS.
    
    In this scenario, the most relevant threats associated with the attack goals are \textit{Spoofing} (e.g., The attacker might impersonate another SIP user by manipulating SIP headers), \textit{Tampering} (e.g., the attacker modifies the Session Description Protocol to change the media streams or redirect calls), \textit{Information Disclosure} (e.g., intercepting sensitive information within the SIP headers or message bodies), and \textit{Elevation of privileges} (e.g., manipulating the SIP communication in such a way that they gain access to resources or actions beyond what they are authorized to).
   
    \item {\em Scenario 2:} An external attacker trying to exploit a single function in the IMS network (e.g., edge functions/devices). One example is, an attacker compromising vulnerability (CVE-2019-15107 \cite{cve2019-15107}) found on signaling devices (e.g., P-CSCF) to disrupt service or a denial-of-service attack against the system. The vulnerability CVE-2019-15107 primarily poses risks related to Denial of Service, but a little effect related to Spoofing and Tampering. In this scenario, we consider single systems as well, especially the edge devices.

\end{itemize}

\subsection{Threats Evaluation at Different Levels}
\label{sec:ThreatEvaluation}
The network, attack model, and method of collecting vulnerability information are described in section \ref{sec:net_attacker_models}. This information is used in the simulations, including the network in Figure~\ref{fig:network} and vulnerability information shown in Table \ref{tbl:host_stride_weightmetrics}. The vulnerabilities used were collected from the National Vulnerability Database \cite{nvd}. 
The focus of the simulation is to demonstrate the threat-specific risk analysis taking into account single or multiple threats to understand the security posture using the IMS-HAG model.

\begin{table}[!htb]
\centering
\tiny
\caption{vulnerability: cvss impact scores, exploitability score, and weights of threats based on nvd data and identified threats}
\label{tbl:host_stride_weightmetrics}
\resizebox{0.54\textwidth}{!}{
\begin{tabular}{|l|l|c|c|c|c|c|c|c|c|}
\hline
\textbf{systems} & \textbf{vuls ID} & \textbf{$aim_{v_i}$} & \textbf{ES} & \textbf{S} & \textbf{T} & \textbf{R} & \textbf{I} & \textbf{D} & \textbf{E} \\ \hline
P-CSCF & CVE-2019-15107 & 5.9 & 3.9 & 0.15 & 0.15 & 0.00 & 0.00 & 0.70 & 0.00 \\ \hline
I-CSCF & CVE-2018-7285 & 3.6 & 3.9 & 0.00 & 0.00 & 0.00 & 0.00 & 1.00 & 0.00 \\ \hline
S-CSCF & CVE-2021-21366 & 1.4 & 2.8 & 0.00 & 0.00 & 0.00 & 0.50 & 0.00 & 0.50 \\ \hline
BCGF & CVE-2019-5437 & 1.4 & 3.9 & 0.00 & 0.40 & 0.00 & 0.60 & 0.00 & 0.00 \\ \hline
MGCF & CVE-2018-5381 & 3.6 & 3.9 & 0.20 & 0.40 & 0.00 & 0.40 & 0.00 & 0.00 \\ \hline
SGW & CVE-2018-5392 & 6.4 & 3.9 & 0.20 & 0.40 & 0.00 & 0.40 & 0.00 & 0.00 \\ \hline
SIP-AS & CVE-2018-10544 & 5.9 & 3.9 & 0.10 & 0.20 & 0.00 & 0.20 & 0.00 & 0.50 \\ \hline
MGW & CVE-2018-5390 & 3.6 & 3.9 & 0.00 & 0.00 & 0.00 & 0.00 & 1.00 & 0.00 \\ \hline
OSA-SCS & CVE-2016-9905 & 5.9 & 2.8 & 0.00 & 0.40 & 0.00 & 0.00 & 0.20 & 0.40 \\ \hline
IM-SSF & CVE-2017-3849 & 4.0 & 2.8 & 0.35 & 0.10 & 0.00 & 0.00 & 0.10 & 0.40 \\ \hline
MRFC & CVE-2022-20053 & 5.9 & 1.8 & 0.00 & 0.00 & 0.00 & 0.00 & 1.00 & 0.00 \\ \hline
MRFP & CVE-2023-49699 & 5.9 & 1.8 & 0.25 & 0.25 & 0.00 & 0.25 & 0.00 & 0.25 \\ \hline
\end{tabular}
} 
\end{table}

\subsubsection{Attack Scenario 1: Modeling and evaluation}
\label{sec:attack_scenario1}

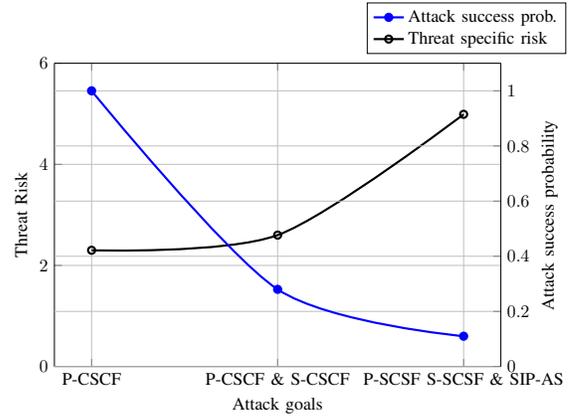
\begin{figure}[!h]
   \centering
\resizebox{0.42\textwidth}{!}{
\begin{tikzpicture}
\pgfplotsset{
    scale only axis,
    symbolic x coords={P-CSCF, P-CSCF \& S-CSCF, P-SCSF S-SCSF \& SIP-AS},
     xtick=data,
     grid=both,
     x=3.7cm, 
     legend pos= north west,
        legend style={cells={anchor=west}, at={(0.7,1.2)}}, anchor=north, 
      every axis plot/.append style={line width=1.2pt},
}
\begin{axis}[
  axis y line*=left,
   ymin=0, ymax=6,
  xlabel=Attack goals,
    ylabel=Threat Risk,
]

\addplot[smooth,mark=o,black]
  coordinates{ 
    (P-CSCF,2.30)
    (P-CSCF \& S-CSCF,2.60)
    (P-SCSF S-SCSF \& SIP-AS,4.99)
}; \label{plot_1}
\end{axis}
\begin{axis}[
  axis y line*=right,
  axis x line=none,
  ymin=0.0, ymax=1.1,
  ylabel=Attack success probability
]
\addplot[smooth,mark=*,blue]
  coordinates{ 
    (P-CSCF,1.0)
    (P-CSCF \& S-CSCF, 0.28)
    (P-SCSF S-SCSF \& SIP-AS, 0.11)
}; \label{plot_2}
\addlegendimage{/pgfplots/refstyle=plot_1}\addlegendentry{Attack success prob.}
\addlegendimage{/pgfplots/refstyle=plot_2}\addlegendentry{Threat specific risk}
\end{axis}
\end{tikzpicture}
}
 \caption{Scenario 1 - Evaluating multiple targets as attack goals}
   \label{fig:multiplegoals}
\end{figure}

In attack scenario 1, we analyze how security metrics change when more than one attack goals are targeted, taking into account all the threats to the attack goals. For this scenario, we conducted three different simulation runs. In the first simulation run, only P-CSCF is considered as the attacker’s goal. In the second simulation run, the goals are expanded to include S-CSCF, and in the third run, the SIP-AS system is added as an additional goal. In this evaluation, based on the vulnerabilities description on the NVD database, the relevant sub-threats identified are—$\{S, T, I, E\}$ for P-CSCF, —$\{I, E\}$ for S-CSCF and —$\{S, T, I, E\}$ for SIP-AS. More details are provided in Table \ref{tbl:host_stride_weightmetrics}. However, in the simulation, we compute threat-specific risk, taking into account all the threats. Here, our focus is to assess and analyze the effect of multiple attack goals in terms of all threats-risk, and the probability of attack success. The results are shown in Figure~\ref{fig:multiplegoals}.

\vspace{5mm} 

\begin{figure*}[tb]
    \centering
    \resizebox{0.32\textwidth}{!}{
    \subfigure[The attack success probability values at the different levels]{
        \label{fig:prob-analysis}
        \begin{tikzpicture}
        \begin{axis}[
            ybar,
            bar width=10,
             scale only axis,
            enlargelimits=0.1,
              legend columns=3, legend style={cells={anchor=west}, at={(0.5,1.1)}, anchor=north, draw=none, name=legend_name,draw} ,
            ylabel={Attack Success Probability},
            xlabel=Attack goals,
            symbolic x coords={P-CSCF, S-CSCF,SIP-AS,MRFP, MGW},
            xtick=data,
            ymin= 0.0,ymax=1,
            width=0.5\textwidth,
            height=6.5cm,
            grid=both,
            nodes near coords align={vertical},
            ]
            \addplot coordinates {(P-CSCF,0.39) (S-CSCF,0.39) (SIP-AS,0.39) (MRFP,0.18) (MGW,0.39) };
            \addplot coordinates {(P-CSCF,0.39) (S-CSCF,0.39) (SIP-AS,0.39) (MRFP,0.18) (MGW,0.39) };
            \addplot coordinates {(P-CSCF,1) (S-CSCF,0.28) (SIP-AS,0.11) (MRFP,0.05) (MGW,0.006) };
            \legend{node level, attack path (max),network level}
            \end{axis}
            \end{tikzpicture}
}}
    \resizebox{.32\textwidth}{!}{
    \subfigure[Individual threats: Threats specific risk with respect to one attack goal (SIP-AS))]{
        \begin{tikzpicture}
        \begin{axis}[
            ybar,
            bar width=10,
            enlargelimits=0.1,
            scale only axis,
              legend columns=3, legend style={cells={anchor=west}, at={(0.5,1.1)}, anchor=north, draw=none, name=legend_name,draw} ,
            ylabel={Threat risk (specific)},
            xlabel=Individual threat only,
            symbolic x coords={S,T,R,I,D,E},
            xtick=data,
            width=0.5\textwidth,
            height=6cm,
            grid=both,
            nodes near coords,
            nodes near coords style={font=\tiny}
            ]
            \addplot coordinates {(S,0.23) (T,0.5) (R,0)(I,0.5) (D,0) (E, 1.2)};
            \addplot coordinates {(S,.6) (T,0.8) (R,0) (I,0.7) (D,1.6) (E,1.4)};
            \addplot coordinates {(S,2.1) (T,4.0) (R,0) (I,1.3) (D,7.5) (E,2.8)};
            \legend{node level, attack path(max),network level}
            \end{axis}
            \end{tikzpicture}
}}
    \resizebox{.32\textwidth}{!}{
    \subfigure[Sub-threats: Threats specific risk values at the different levels (for $\{S,T,E\}$)]{
        \label{fig:threats-risk}
        \begin{tikzpicture}
        \begin{axis}[
            ybar,
            bar width=10,
            enlargelimits=0.12,
             scale only axis,
              legend columns=3, legend style={cells={anchor=west}, at={(0.5,1.1)}, anchor=north, draw=none, name=legend_name,draw} ,
            ylabel={Threat risk (specific)},
            xlabel=Attack goals,
            symbolic x coords={P-CSCF, S-CSCF, SIP-AS, MRFP, MGW},
            xtick=data,
            width=0.5\textwidth,
            height=6.2cm,
            grid=both,
            nodes near coords,
            nodes near coords style={font=\tiny}
            ]
            \addplot coordinates {(P-CSCF,0.7) (S-CSCF,0.2) (SIP-AS,1.8) (MRFP,0)(MGW,0) };
            \addplot coordinates {(P-CSCF,0.7) (S-CSCF,0.9) (SIP-AS,2.7) (MRFP,0.90) (MGW,3.4) };
            \addplot coordinates {(P-CSCF,0.7) (S-CSCF,1.8) (SIP-AS,5.5) (MRFP,3.7) (MGW,6.9) };
            \legend{node level, attack path(max),network level}
            \end{axis}
            \end{tikzpicture}
}}
\caption{Scenario 2 - Evaluating attack impact based on different attack goals and STRIDE Model using IMS-HAG}
    \label{fig:Comaprison-ImpactattackGoals}
\end{figure*}
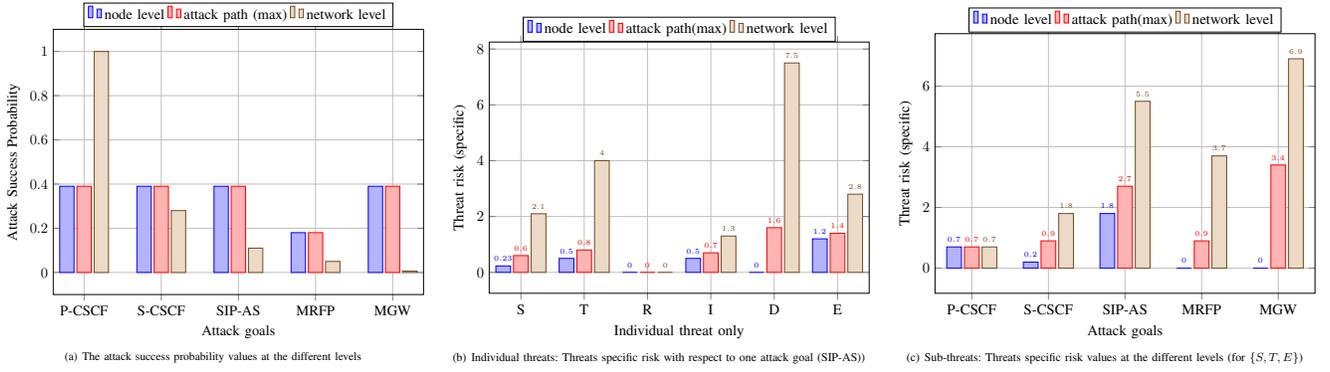

The results illustrate how the attack success probability and threat-specific metrics change when multiple targets are taken into account. As in the results, as more attack goals are included in the evaluation, the threat-specific metrics show an increase when a new goal is added. This trend shows that the impact of threats and attacks on the systems rises when an attacker targets multiple systems. Conversely, the attack success probability decreases from 1.0 to 0.11 as new goals are added. This decrease signifies that the likelihood of an attacker successfully compromising multiple goals is lower than when targeting a single goal.

\subsubsection{Security Evaluation: Attack Scenario 2}
\label{sec:attack_scenario2}

In Figure~\ref{fig:Comaprison-ImpactattackGoals}, we present the security assessment for attack scenario 2, showing the impact of various attacks using sub-threat-specific metrics. Additionally, we illustrate the impact of individual threats at different levels; node level, attack-path level, and network level.

Figure~\ref{fig:prob-analysis} shows the results of the attack success probability (i.e., $p(n_j)$, $p(path_i)$, and $Pr$ levels) for various attack goals, indicating the likelihood of an attacker successfully exploiting vulnerabilities to compromise the respective targets. As expected, the security evaluation results confirm that the attack success probability for MGW is lower than the other goals at the network level due to the multiple nodes required to reach MGW. In contrast, the probability is higher for P-CSCF, as it serves as the initial node and entry point for the attacker at the network level.

We then assess the impact of individual threats on the IMS network, focusing on one function as the attack goal: SIP-AS (the same analysis can be applied to other functions). Figure~\ref{fig:threat_subsystems} shows the potential impact of individual threats in achieving the final goal. At the node level, the metric represents the impact on SIP-AS based on each threat's risk. At the attack-path level, it shows the maximum impact of the threat across all attack paths leading to SIP-AS. The network-level metric evaluates the overall impact of the threat across all attack paths leading to the final goal.

The results indicate that the \textit{'D'} threat has the greatest impact compared to other threats when targeting SIP-AS, with a risk value of 7.5 at the network level. At the path level, the risk value is also highest, reaching 1.6, while the node-level value for \textit{'D'} for the host SIP-AS is zero. This suggests that, although the direct threat to SIP-AS is negligible, the cumulative impact of \textit{'D'} on nodes along the paths to SIP-AS is significant.

We also assess and compare the effects of sub-threats on the IMS network for various attack goals. In this analysis, we focus on \textit{S}, \textit{T}, and \textit{E} as sub-threats. These specific threats were chosen due to their critical impact on system security, though any combination of threats can be selected based on the security manager’s priorities. Figure~\ref{fig:threats-risk} illustrates the potential impact of these sub-threats on the IMS system across different levels.

The results indicate that MGW has the highest threat impact values for \textit{S}, \textit{T}, and \textit{E} at both the network and path levels, while at the node level, the value for these threats is zero. This suggests that although the individual node-level impact is zero, the cumulative threat impact at the path and network levels is significant. In contrast, P-CSCF shows a consistent threat-risk value of 0.7 across all levels for the identified threats. This is because P-CSCF has only one vulnerability, and the attack tree reflects the risk associated with that single vulnerability. Since P-CSCF is the entry point to the network and there are no additional nodes on the attack paths beyond it, the threat-risk values for the node, path, and network levels are identical.

\subsection{Threats Security Evaluation for Sub-systems}
\label{sec:scenario3}
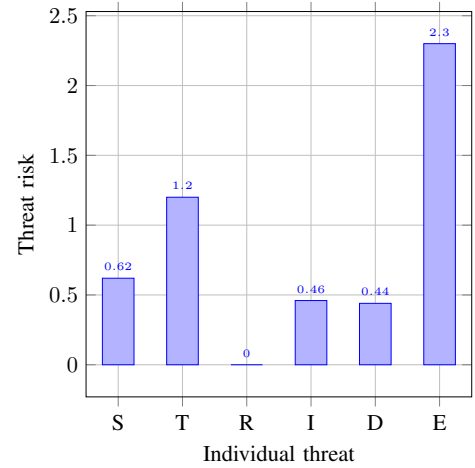
\begin{figure}[!h]
\label{fig_subsystemthreats}
    \centering
    \resizebox{.35\textwidth}{!}{
        \begin{tikzpicture}
        \begin{axis}[
            ybar,
            bar width=14,
            enlargelimits=0.1,
            scale only axis,
              legend columns=3, legend style={cells={anchor=west}, at={(0.5,1.1)}, anchor=north, draw=none, name=legend_name,draw} ,
            ylabel={Threat risk},
            xlabel=Individual threat,
            symbolic x coords={S,T,R,I,D,E},
            xtick=data,
            width=0.5\textwidth,
            height=6cm,
            grid=both,
             x=1.0cm, 
            nodes near coords,
            nodes near coords style={font=\tiny}
            ]
            \addplot coordinates {(S,0.62) (T,1.20) (R,0)(I,0.46) (D,0.44) (E, 2.3)};
            \end{axis}
            \end{tikzpicture}
}
\caption{Evaluation of individual threats based on all systems in the application servers group only; SIP-AS, OSA-SCS, and IM-SSF}
    \label{fig:threat_subsystems}
\end{figure}

In this section, we assess the threats to subsystems (e.g., all the Call Session Control Functions, Application server functions, etc) within the IMS network. Specifically,  we assess the threats of systems located in the application servers group. While we could analyze other groups as well, due to space limitations, we focus only on the application servers in this section. The assessment covers the impact of threats on systems such as SIP-AS, OSA-SCS, and IM-SSF, using the \textit{STRIDE} model, as shown in Figure \ref{fig:threat_subsystems}.

\vspace{2mm} 

The results show that \textit{E} has the highest impact compared to other threats. This is because each system within the application servers group weights at least 0.4, whereas other threats either have less weight or a weight of zero. In contrast, \textit{R} shows the lowest impact, as all threats have zero weights for repudiation.

\subsection{Evaluating the Effectiveness of Defense Measures using IMS-HAG}
\label{sec:sdefense_measures}
\begin{figure}[!h]
 	\centering
 \includegraphics[width=1.0 \linewidth]{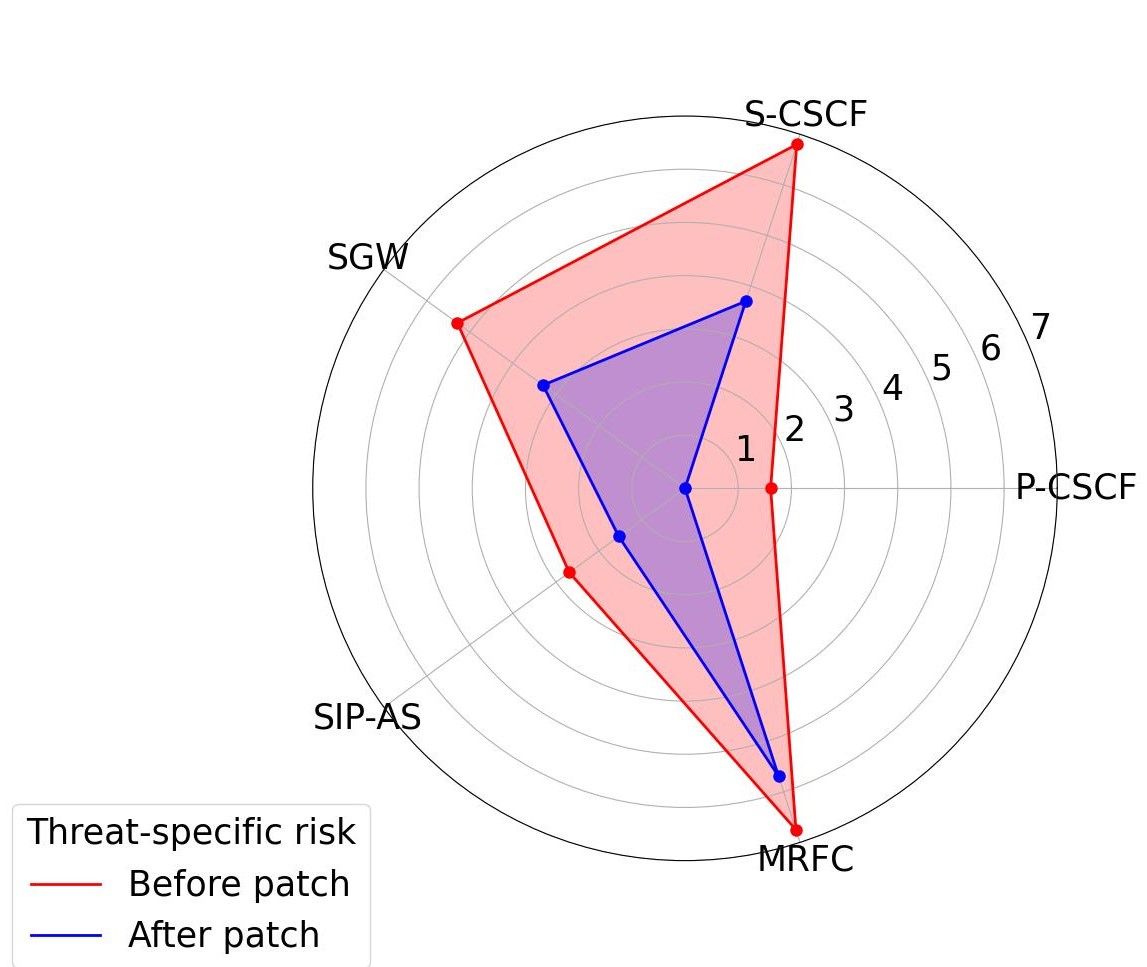}
 	\caption{Defense Evaluation using threat-specific risk based on different threats related to the respective functions}
 	\label{fig:radar}
 \end{figure}

It is important to evaluate the effectiveness of defense mechanisms before deploying them on IMS networks, as different functions within the IMS have unique security requirements. For example, the P-CSCF may face various threats, but Denial of Service (DoS) is especially the most critical due to its role in managing signaling traffic. So, this assessment will specifically target the threats - DoS for the P-CSCF, information disclosure and elevation of privilege for S-CSCF, information disclosure for BCGF, and tampering and information disclosure for SGW. While a combination of other threats can be considered, we will demonstrate only a few due to page number limits. The results for the threat-specific risk are shown in Figure \ref{fig:radar}.

 \vspace{5mm} 

The results indicate that deploying the patch on the P-CSCF reduced the threat risk for `D' to zero. This is because the P-CSCF serves as the entry point to the network and eliminating all vulnerabilities on it has improved the threat security of the system for the DoS. P-CSCF is the first function encountered when accessing the network, and no other functions need to be compromised to reach it, so no risk related to other functions at this point.

In contrast, for other functions that require compromising the P-CSCF and S-CSCF before they can be accessed, the threat-specific risk did not reduce completely to zero but has decreased compared to the ``before patch" state. This is due to the presence of other threats associated with the systems along the paths leading to these targets, which means the overall threat risk could not be completely removed. In this case, threats on systems on the attack paths to the final target will need to be addressed if the risk is to be reduced.

Additionally, the MRFC showed only a slight reduction in risk because it can be reached through two different attack paths, each presenting similar threats, even after the host-based threat was patched.

\section{Conclusions} 
\label{sec:conclusion}

Existing studies have shown a lack of capability to comprehensively capture and assess the threats to IMS networks, considering both the threats and risks. In this paper, we propose a hierarchical security model for IMS networks aimed at evaluating vulnerabilities and threat-specific risks with the STRIDE model threat categories. We present the formalism of the model and metrics (node, path, and network levels). We perform a preliminary evaluation based on vulnerabilities collected from the National Vulnerability Database for devices in the IMS network.
Furthermore, we demonstrate the model's applicability and present the results of implementations for three different attack scenarios. Before deploying defense mechanisms on IMS networks, we assess their effectiveness, considering the unique security requirements of various IMS functions, and also evaluate the threat-specific risks after implementing each defense mechanism.
The results show that we can model and analyze the threats to IMS networks using the proposed model and compute threat-specific risks for systems in the IMS network. We believe that network defenders can use this model to understand their security posture, taking into account the threats and risks posed by each vulnerability in the IMS network.

\addcontentsline{toc}{section}{References}
\bibliographystyle{IEEEtran}
\bibliography{mybibfile}
\end{document}